\documentclass[prd,onecolumn]{revtex4}
\pdfoutput=1
\usepackage{amssymb,latexsym}
\usepackage{amsmath,amsbsy,bbm}
\usepackage{ifpdf}
\usepackage{epsfig,bm}
\usepackage{graphicx,comment}
\usepackage{color}
\usepackage{mathtools}
\usepackage{comment}
\usepackage[normalem]{ulem}
\begin{document}

\title{Berezinskii--Kosterlitz--Thouless transition of the two-dimensional $XY$ model on the honeycomb lattice}

\author{Fu-Jiun Jiang}
\email[]{fjjiang@ntnu.edu.tw}
\affiliation{Department of Physics, National Taiwan Normal University,
88, Sec.4, Ting-Chou Rd., Taipei 116, Taiwan}

\begin{abstract}
  The Berezinskii--Kosterlitz--Thouless (BKT) transition of the two-dimensional
  $XY$ model on the honeycomb lattice is investigated using Neural Network (NN) and Monte Carlo simulations. It is demonstrated in the literature
  that with certain plausible assumptions, the associated critical
  temperature $T_{\text{BKT,H}}$ is found to be $\frac{1}{\sqrt{2}}$ exactly. Surprisingly,
  the value of $T_{\text{BKT,H}}$ obtained from our NN calculations is 0.560(9) which
  deviates significantly from $\frac{1}{\sqrt{2}}$. In addition, based on the
  helicity modulus,
  the $T_{\text{BKT,H}}$ determined is 0.571(8) agreeing well with that resulting from the
  NN estimation. The outcomes presented in this study indicate that a
  detailed analytic calculation is desirable to solve the found discrepancy. 
  
\end{abstract}

\maketitle

\section{Introduction}

The two-dimensional (2D) $XY$ model has been one of the research topics in phase transitions. In particular, instead of normal second-order phase
transition which is related to spontaneous symmetry breaking, the 2D $XY$ model
has a Berezinskii--Kosterlitz--Thouless type
transition which is associated with topological defects \cite{Ber71,Ber72}. Specifically,
as the temperature increases, the model exhibits a transition from a phase
of bound vortex-antivortex pairs to a phase having unbound
vortices and antivortices \cite{Kos72,Kos73,Kos74}. Such a transition is of infinite order
in Ehrenfest’s scheme.
 
Due to its unique phase transition as well as the relevance to experiments \cite{Bis78,Eps81,Hu20},
the 2D $XY$ model has been studied extensively \cite{Jos13}. In particular,
the properties of this model on the square lattice have been determined with high precision.
For example, the inverse transition temperature $\beta_{\text{BKT,S}}$ of the 2D $XY$ model on the square
lattice is calculated accurately to be 1.1199(1) \cite{Has97,Has05}.

During the last decade, one has witnessed an era of applications of Machine Learning (ML) methods
in various research fields \cite{Baldi:2014pta,Mnih:2015jgp,Hoyle:2015yha,For23,Aza23}. Moreover, techniques of Neural Networks (NN) have been adopted in uncovering
various phases of matters. In particular, NNs are shown to be able to calculate the critical points of many
physical models \cite{Car16,Nie16,Den17,Li18,Chn18,Rod19,Zha19,Tan20.1}. It is worth mentioning that a simple multilayer perceptron (MLP)
which has only one hidden layer of two neurons is proved to be universal \cite{Pen22,Tse23,Tse231}, namely the same MLP, without
conducting any new training, has
been employed to determine the critical temperatures of many three-dimensional (3D) and two-dimensional (2D)
models successfully.
Apart from this, it is implied in several studies that NN-related
approaches may speed up the calculations significantly \cite{Hua17,She18,Paw20,Sar21,Tha22}. For instance, it is demonstrated in Ref.~\cite{Din22}
that
in a Monte Carlo simulation, the configurations obtained much ahead of reaching the equilibrium stage
can be employed to identify the critical points of several models with high precision.

By mapping a 2D $n$-component spin model onto a solid-to-solid model and using
certain plausible assumptions, the critical temperature $T_{\text{BKT,H}}$ of the 2D $XY$ model on the honeycomb
lattice is found to be $1/\sqrt{2}$ in Ref.~\cite{Nie82}. Although
$T_{\text{BKT,H}}$ = $1/\sqrt{2}$ is employed in some later calculations \cite{Den07,Wan21}, it seems that the
result is never confirmed by an exact numerical method. Because of this, here, we apply both the techniques of NN
and Monte Carlo simulation (MC) to determine the critical temperature $T_{\text{BKT,H}}$ of
the 2D $XY$ model on the honeycomb lattice.

Surprisingly, the NN outcome for the $T_{\text{BKT,H}}$ of the considered model is given by $0.560(9)$.
In addition,
by applying the expected finite-size ansatz to the
relevant Monte Carlo data, we find that $T_{\text{BKT,H}} = 0.571(8)$.
Both the values of $T_{\text{BKT,H}}$, obtained from two different approaches,
differ from $1/\sqrt{2}$ significantly. In particular, we find that
the estimated pseudo-critical temperatures $T_{\text{BKT,H}}(L)$ is drifting away from $1/\sqrt{2}$ with increasing $L$.

The critical temperature $T_{\text{BKT,H}}$ of the 2D classical $XY$ model on the honeycomb lattice is determined to be $T_{\text{BKT,H}} \sim 1.1634$
in Ref.~\cite{Cam96} by a strong coupling analysis of the
two-dimensional $O(N)$ $\sigma$ model. Later we will argue that the calculations done in Ref.~\cite{Cam96} are likely consistent with our results.

In conclusion, the presented results in this investigation indicate that a refinement of
the related analytic calculation is required to better understand the
deviation between the numerical outcomes reached here and the theoretical
prediction of Ref.~\cite{Nie82}.

The rest of the paper is organized as follows. After the introduction, the considered 2D $XY$ model
on the honeycomb lattice as well as the calculated observables are described in Sect.~II. Moreover, in Sect.~III, the employed NN is
summarized. We then present the numerical results from both the NN and the Monte Carlo data
in Sect.~IV. Finally, the conclusions of the present study are given in Sect.~V.

\begin{figure}
  \vskip-0.5cm

       \includegraphics[width=0.2\textwidth]{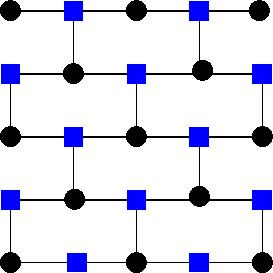}        
        \vskip-0.2cm
        \caption{The periodic 4 by 4 honeycomb lattice implemented in this
          study.}
        \label{honey}
\end{figure}
  
\section{The considered model and observable}

The Hamiltonian of the 2D $XY$ model considered here has the following expression \cite{Has05,Wol89,Bra22}
\begin{equation}
H = -\sum_{\left< ij\right>} \vec{e}_i\cdot\vec{e}_j,
\label{eqn}
\end{equation}
where $\left<ij\right>$ refers to the nearest
neighbor sites $i$ and $j$, and $\vec{e}_i$ is a vector at site $i$ with
$\vec{e}_i \in S^2$.

Fig.~\ref{honey} demonstrates the periodic honeycomb lattice implemented in this
investigation.

The helicity modulus $\Gamma$ is calculated here. This quantity describes how
the free energy reacts when an infinitesimal change of the
boundary conditions is applied to the system. Explicitly, $\Gamma$ is given by \cite{Has05,Bra22}
\begin{eqnarray}
	&&\Gamma =\frac{4}{3\sqrt{3}}\Bigg(\frac{1}{L^2}\left< \sum_{i}^{N} \vec{e}_i\cdot\vec{e}_{i+1}\right> \nonumber \\
	&&-\frac{1}{TL^2}\left<\left(\sum_{i}^{N}\left(e^1_ie^2_{i+1}-e^2_ie^1_{i+1}\right)^2\right)\right>\Bigg).
\end{eqnarray}
Here $T$ and $L$ are the temperatures and the linear system size ($N = L^2$), and we assume that at the boundaries the twist is applied in the $x$-direction.
The factor $\frac{4}{3\sqrt{3}}$ appearing above is the ratio of the spin densities on the honeycomb and the square lattices.

\section{The numerical results}

The Monte Carlo simulations (MC) required to calculate the considered critical point $T_{\text{BKT,H}}$ of the 2D
$XY$ model on the honeycomb lattice is conducted using the Wolff algorithm \cite{Wol891}.

\subsection{The employed MLP}

The NN used here is directly adopted from Refs.~\cite{Tan21,Tse22}. The detailed descriptions regarding the construction of the NN
are outlined in Refs.~\cite{Tan21,Tse22}. Here for completeness, we briefly introduce the NN used in this investigation as well as the associated training procedure.

\begin{figure*}
       \includegraphics[width=0.8\textwidth]{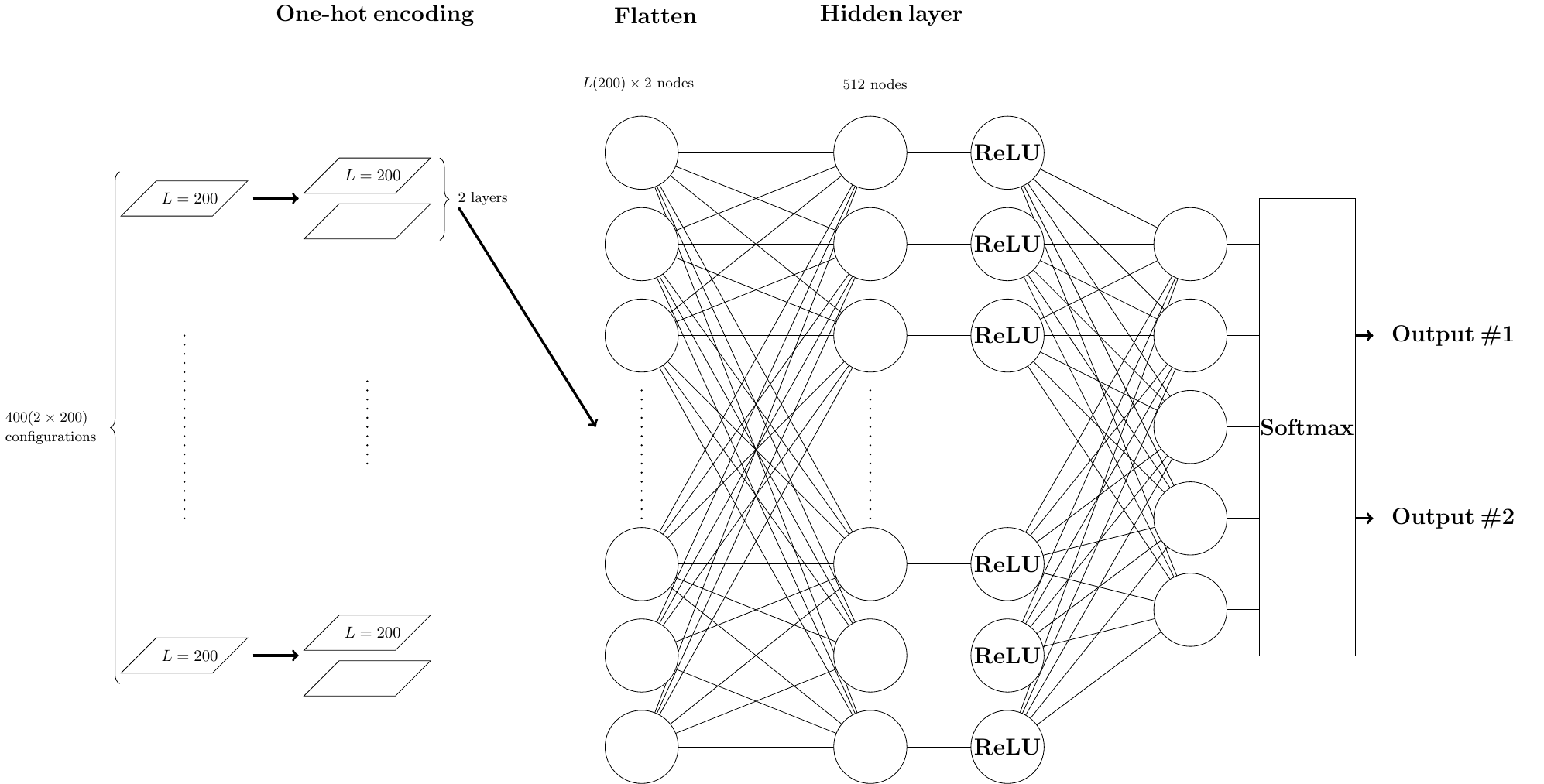}
        \vskip-0.2cm
        \caption{The MLP employed in this study \cite{Tan21,Tse22}.}
        \label{figMLP}
\end{figure*}

The considered NN consists of one input layer, one hidden layer of 512 neurons,
and one output layer (no training is carried out in the present study).
The algorithm employed is minibatch and the optimizer considered is adam. $L_2$ regularization is used in the NN calculations
to avoid overfitting. In the NN architecture, activations functions ReLU and softmax are applied. 800 epochs are performed
and the batch size is 40. Fig.~\ref{figMLP} is the cartoon representation for
the described NN \cite{Tan21,Tse22}.

The training set for the NN consists of 200 copies of two artificially made
configurations. Each of the configurations has 200 elements. In particular,
all the elements of the first and the second configurations have the value of 1 and 0, respectively.
The labels used for these two types of training objects are (1,0) and (0,1). 
It is demonstrated that the NN with such a training strategy is capable of
determining the critical points of several systems \cite{Tan21,Tse22,Tse231}.

Finally, 10 sets of random seeds are used in our calculations. Therefore there will be 10 NN outcomes.
The results presented in the following subsections are based on these 10 NN outcomes.

\subsection{The $T_{\text{BKT,H}}$ of the 2D $XY$ model on the honeycomb lattice determined by the NN method}

In this study, the magnitude of the NN output vectors, denoted by $R$, will be used to
determine the critical temperature $T_{\text{BKT,H}}$ of the considered 2D $XY$ model on the honeycomb lattice.
Since each of the two types of configurations used in the training set has 200
elements, the required configurations for the NN prediction should have
200 spins. Hence, the configurations for the NN prediction are constructed
from the raw spin
configurations through the following steps. First of all, two hundred spins
are chosen randomly. Second, $\phi\,\, \text{mod}\,\, \pi$ of these picked spins are
employed to build the configurations for the NN prediction. A typical configuration after the modulus $\pi$ step has the form (1,1,0,1,0,0,0,1,...,0,0,1).

\begin{figure*}
  \hbox{
    \includegraphics[width=0.5\textwidth]{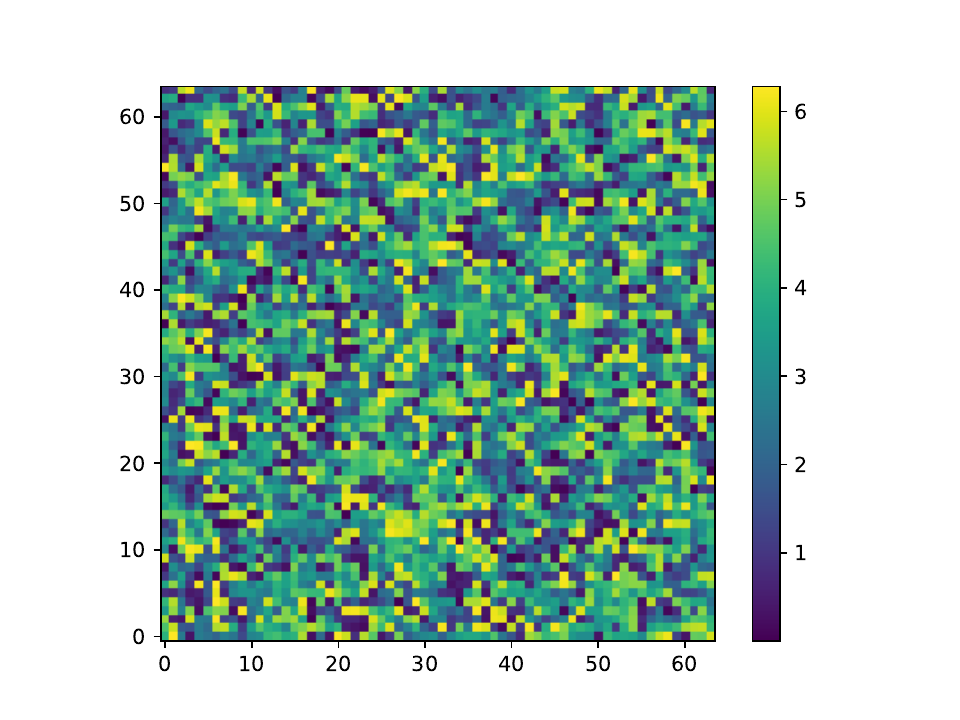}
    \includegraphics[width=0.5\textwidth]{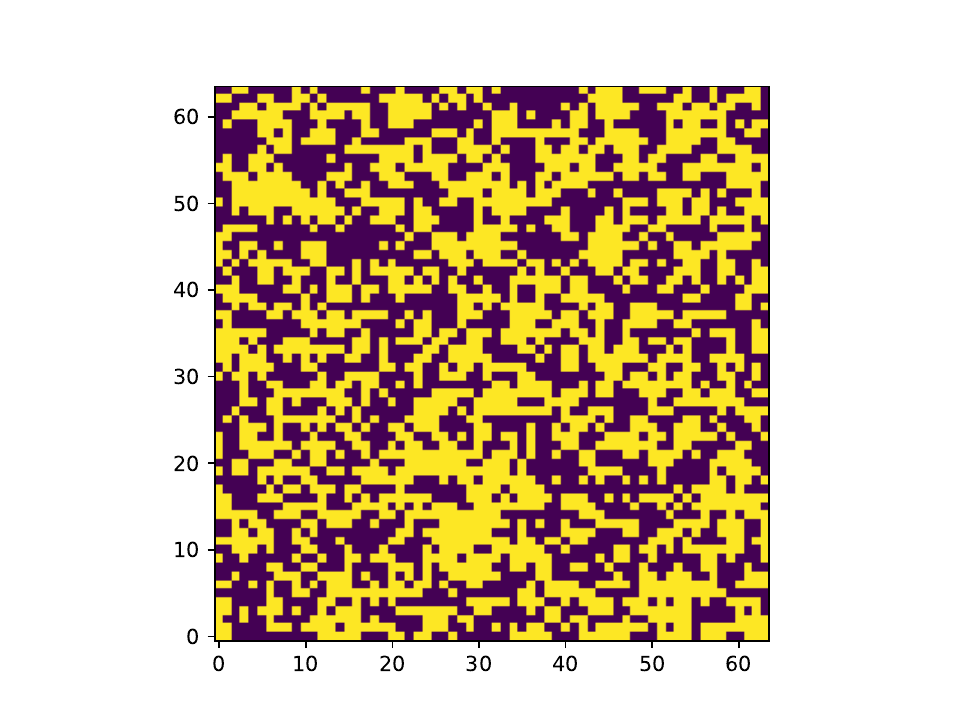}
    }
        \vskip-0.2cm
        \caption{(Left) The snapshot of a configuration obtained at the high-temperature region. (Right) The result
        obtained after the mod $\pi$ procedure for a configuration determined at the high-temperature region.}
        \label{snapshotH}
\end{figure*}

If a configuration is obtained at the high-temperature region,
the vortices and anti-vortices are not bound. In addition,
the distribution of these topological objects is random and has no specific
pattern. As a result, the angles of the spins are random as well. This would
lead to a configuration in which the associated elements are arbitrary in 1 and 0
after the mod $\pi$ step. The output vector of such a configuration is $\sim (0.5,0.5)$
which has $R \sim 1/\sqrt{2}$. The left panel of fig.~\ref{snapshotH} is a snapshot of a configuration obtained at a high temperature.
As can be seen from the figure, the distribution of the angles is quite random. The right panel of fig.~\ref{snapshotH}
shows the result after the mod $\pi$ procedure for a typical configuration at the high-temperature region. The outcome indeed 
demonstrates the arbitrariness of the distribution of 1 and 0.

\begin{figure*}
  \hbox{
    \includegraphics[width=0.5\textwidth]{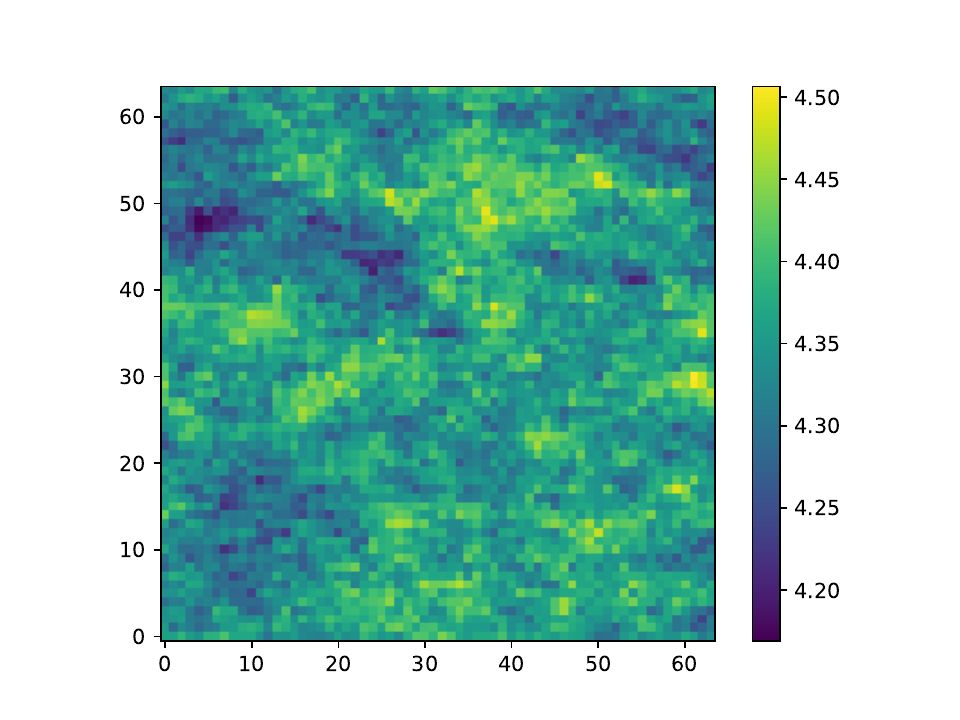}
    \includegraphics[width=0.5\textwidth]{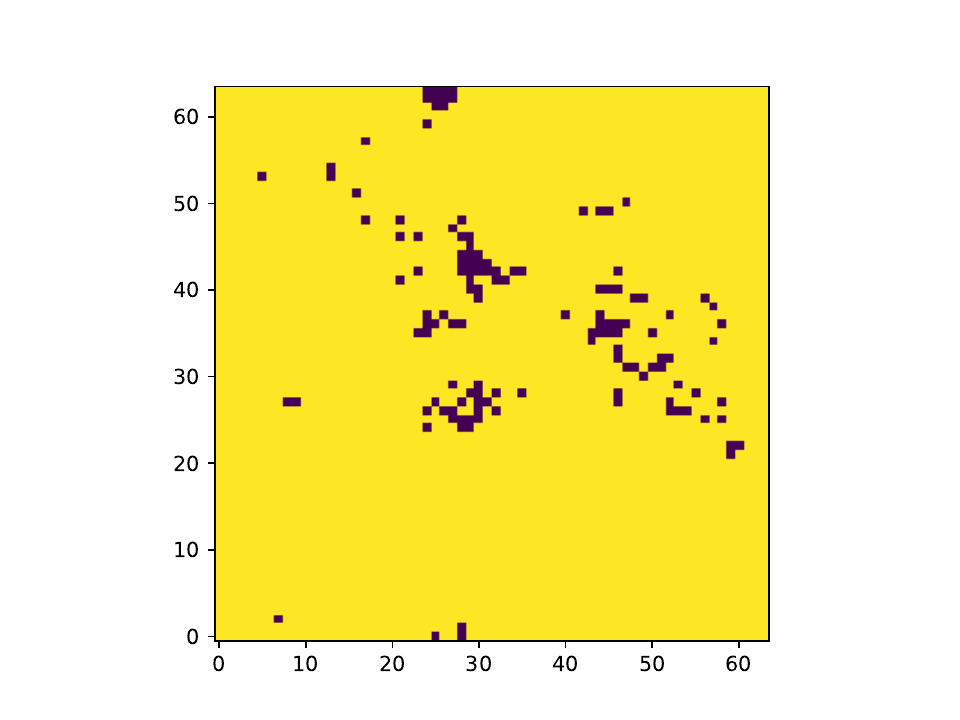}
    }
        \vskip-0.2cm
        \caption{(Left) The snapshot of a configuration obtained at extremely low-temperature region. (Right) The result
        obtained after the mod $\pi$ procedure for a configuration determined at the extremely low-temperature region.}
        \label{snapshotL}
\end{figure*}

Interestingly, at extremely low temperatures, the majority of the raw spin configurations
have one common feature, namely the associated spins' angles
$\theta s'$ satisfy either $\theta s' < \pi$ or $\theta 's > \pi$, see the left panel of fig.~\ref{snapshotL}
for a typical snapshot of this described scenario. The right panel
of fig.~\ref{snapshotL} is a representative snapshot after the modulus $\pi$ step for a configuration in the extremely low-temperature region.
The characteristic described above for configurations in the extremely low-temperature region will lead to output vectors around (1,0) or (0,1) which have $R \sim 1$.
In conclusion, as one goes from the (extremely) low-temperature region to the high-temperature region, the value of $R$ changes from $1$ to a
number close to $1/\sqrt{2}$. As we will demonstrate immediately, this is indeed what's been observed.

\begin{figure*}
  \hbox{
    \includegraphics[width=0.45\textwidth]{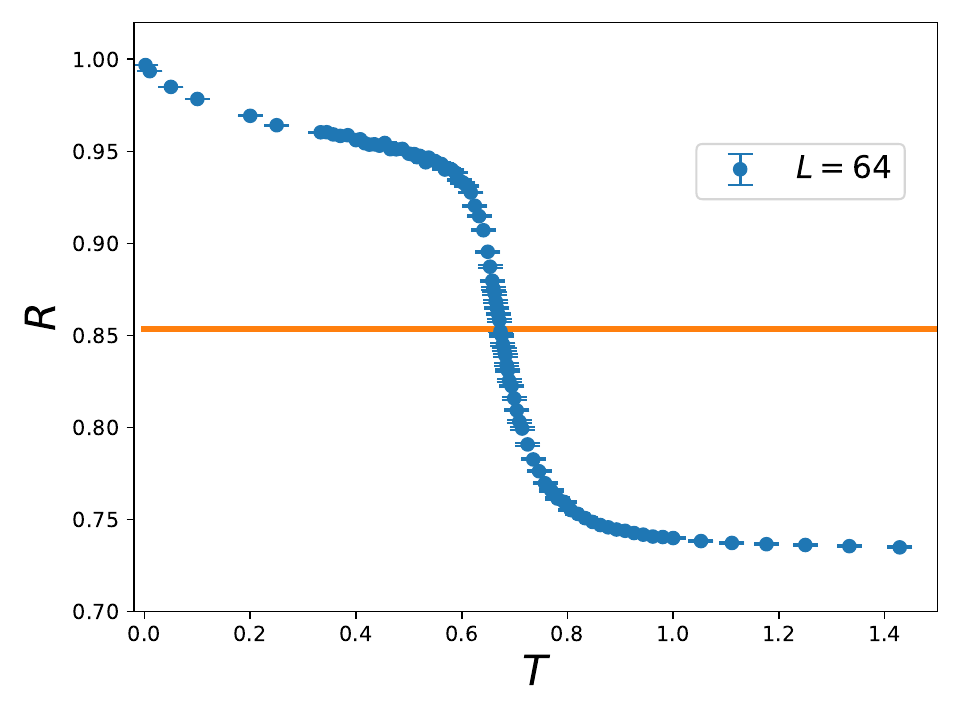}~~~~
    \includegraphics[width=0.45\textwidth]{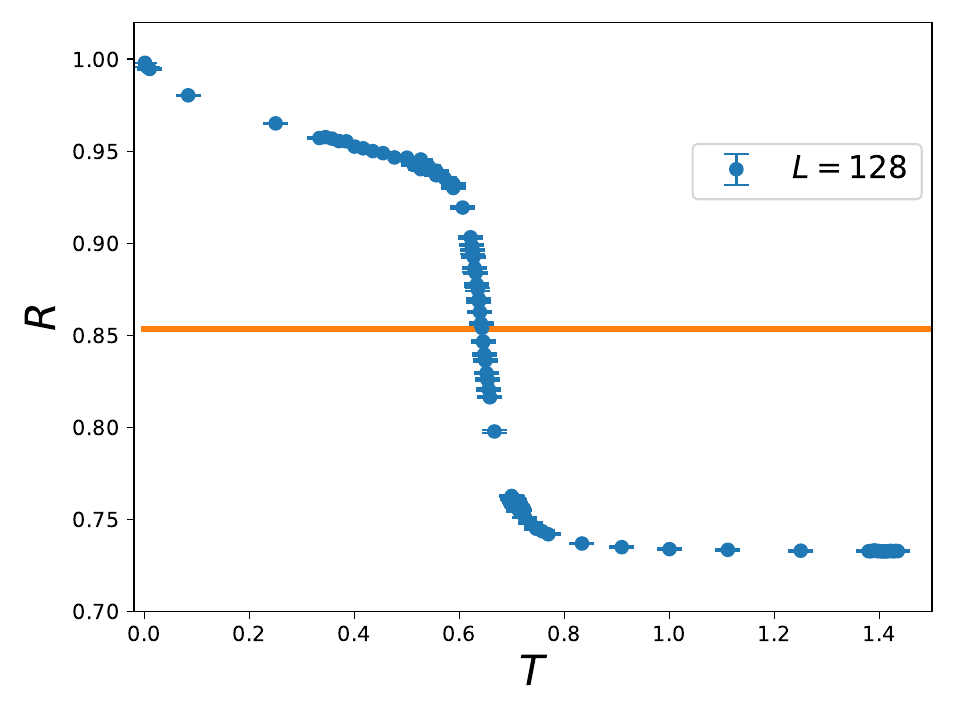}
    }
        \vskip-0.2cm
        \caption{$R$ as functions of $T$ for $L=64$ (left panel) and $L=128$ (right panel). The horizontal solid lines in
        both panels are $(1+1/\sqrt{2})/2$.}
        \label{RL64L128}
\end{figure*}

$R$ as functions of $T$ for $L=64$ and 128 are shown in the left and the right panels of fig.~\ref{RL64L128}, respectively.
Both panels of fig.~\ref{RL64L128} indicate that as one goes from the low-temperature region to the high-temperature region,
the value of $R$ changes from 1 to a number close to $1/\sqrt{2}$.

To calculate the critical temperature $T_{\text{BKT,H}}$, we employ the method used in Ref.~\cite{Tse231}. Specifically, for a given $L$, when the associated $R$
is considered as a function of $T$,
the temperature corresponding to the intersection of $R$ and $(1+1/\sqrt{2})/2$ is taken as the pseudo-critical
temperature $T_{\text{BKT,H}}(L)$ of the given $L$. The horizontal lines in both panels of fig.~\ref{RL64L128} are $(1+1/\sqrt{2})/2$.

\begin{figure*}
    \includegraphics[width=0.6\textwidth]{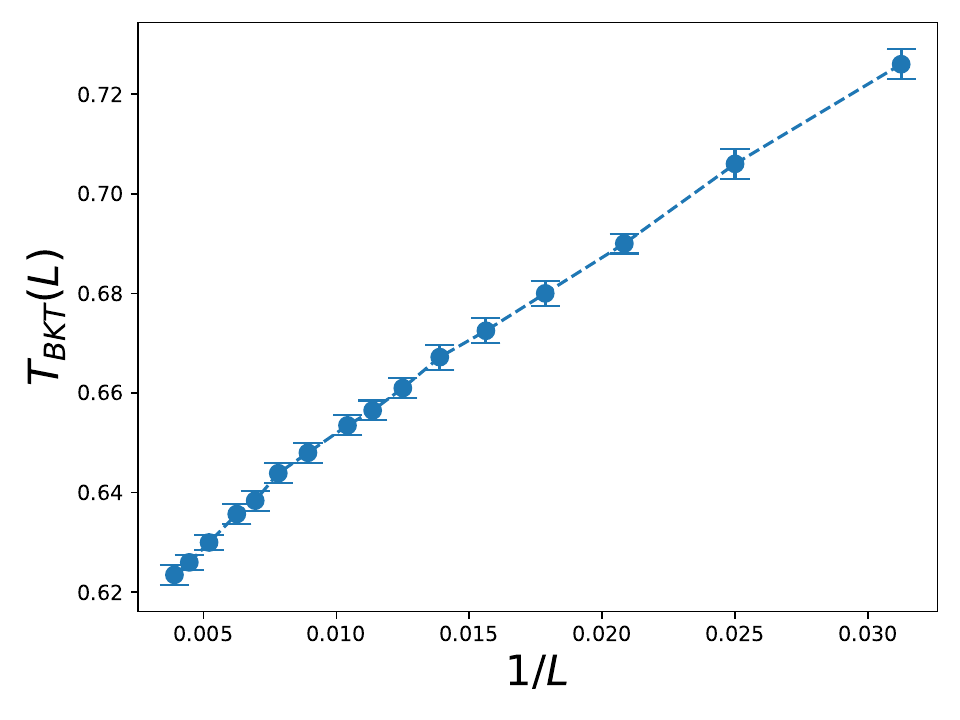}
        \vskip-0.2cm
        \caption{$T_{\text{BKT,H}}(L)$ as a function of $1/L$. The results are obtained from NN approach.}
        \label{TcLL}
\end{figure*}

With the procedure introduced in the previous paragraph, the $T_{\text{BKT,H}}(L)$ as a function of $1/L$ is shown in fig.~\ref{TcLL}.

Theoretically, it is known that the pseudo-critical temperatures on finite lattices satisfy the following ansatz \cite{Nel77,Pal02}
\begin{equation}
  \label{fit}
T_{\text{BKT,H}}(L) = T_{\text{BKT,H}} + a \frac{T_{\text{BKT,H}}}{\left(\log\left(L\right)+c\right)^2}, 
\end{equation}
where in equation (\ref{fit}) $a$ and $c$ are some constants and $T_{\text{BKT,H}}$ is the bulk critical temperature.
A fit of the data in fig.~\ref{TcLL} to the above ansatz leads to
$T_{\text{BKT,H}} = 0.560(9)$. The obtained $T_{\text{BKT,H}}=0.560(9)$ deviates significantly from the theoretical prediction $1/\sqrt{2} \sim 0.70711$.
As we will show in the next subsection, the $T_{\text{BKT,H}}$ obtained by the NN method agrees well with the one determined from Monte Carlo simulations.

\subsection{The $T_{\text{BKT,H}}$ obtained from the helicity modulus $\Gamma$}

\begin{figure}
       \includegraphics[width=0.6\textwidth]{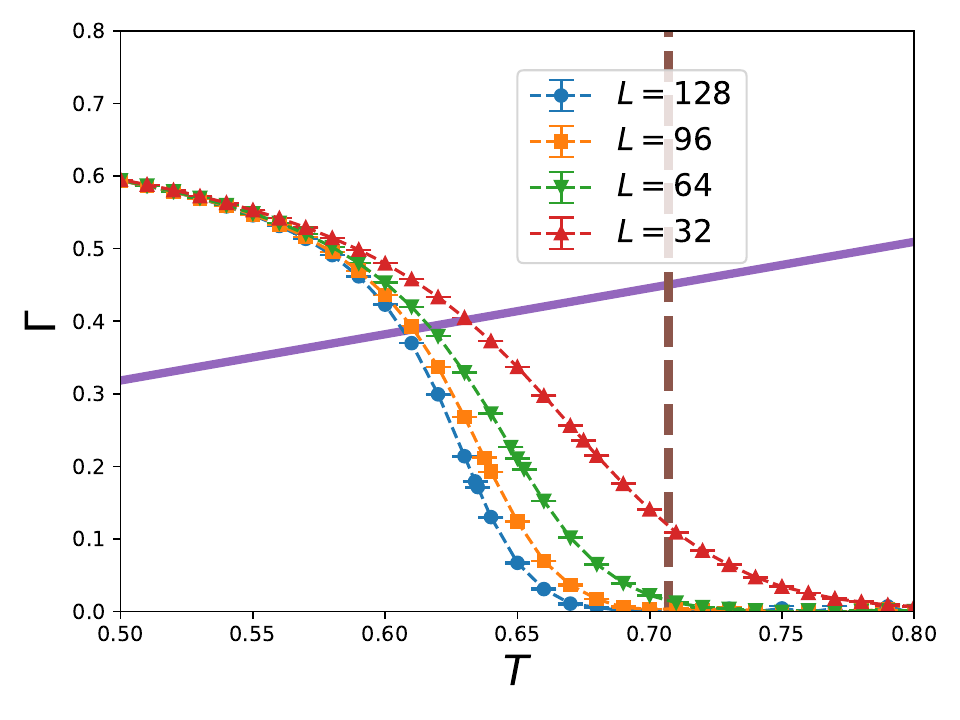}        
        \vskip-0.2cm
        \caption{Intersection points of $2T/\pi$ (solid line) and $\Gamma$ of
          various linear box sizes $L$. The vertical dashed line is
          $1/\sqrt{2}$ (The predicted critical point from Ref.~\cite{Nie82}).}
        \label{gamma}
\end{figure}

Fig.~\ref{gamma} shows the helicity modulus $\Gamma$ of several linear system sizes ($L = 32, 64, 96, 128$)
as functions of $T$. In the figure, the vertical dashed line is $1/\sqrt{2}$ which is the prediction
of $T_{\text{BKT,H}}$ from Ref.~\cite{Nie82} with certain plausible assumptions. In addition,
the solid line in fig.~\ref{gamma} is $2T/\pi$. For each used $L$, the corresponding $T_{\text{BKT,H}}(L)$ is
estimated by the intersection of the associated $\Gamma(L)$ and $2T/\pi$.
It is clear that as $L$ increases, the intersection of $\Gamma$ and $2T/\pi$, namely $T_{\text{BKT,H}}(L)$
drifts away from $1/\sqrt{2} \sim 0.70711$.

\begin{figure}
       \includegraphics[width=0.6\textwidth]{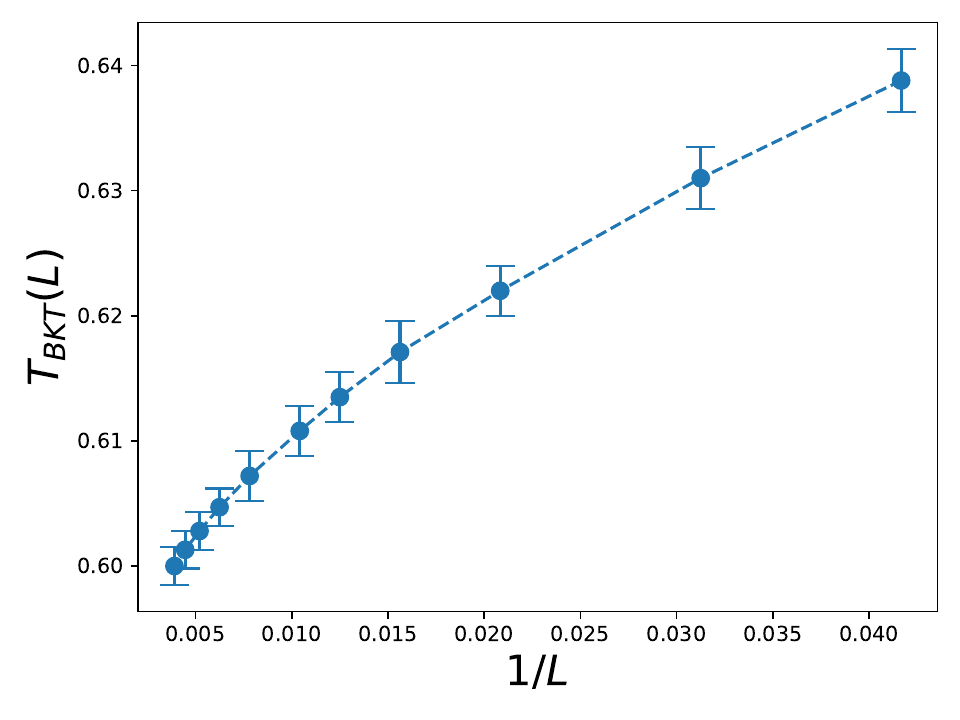}        
        \vskip-0.2cm
        \caption{Pseudo-critical temperatures $T_{\text{BKT,H}}(L)$ as a function
        of the linear box sizes $L$. The results are obtained from $\Gamma$.}
        \label{gamma1}
\end{figure}

$T_{\text{BKT,H}}(L)$ obtained by above described procedure as a function of $1/L$ is demonstrated in
fig.~\ref{gamma1}. A fit using the data of fig.~\ref{gamma1} and ansatz~(\ref{fit})
leads to $T_{\text{BKT,H}} = 0.571(8)$. In performing the fits, certain reasonable constraints are applied. For instance,
only the fits with the results of $T_{\text{BKT,H}} < 1.0$, $c > 0$, and $|a| < 10$ are accepted.  
It is obvious that
the determined $T_{\text{BKT,H}} = 0.571(8)$ agrees with the one obtained by the NN method but differs statistically from
the theoretical prediction $1/\sqrt{2} \sim 0.70711$.

\section{Discussions and Conclusions}

Using the NN approach and MC simulations, we calculate the critical temperature $T_{\text{BKT,H}}$ associated with the BKT transition
for the 2D $XY$ model on the honeycomb lattice.
The obtained outcome $T_{\text{BKT,H}} = 0.560(9)$ and $T_{\text{BKT,H}} = 0.571(8)$ differ significantly from the theoretical
prediction $1/\sqrt{2} \sim 0.70711$.

In Ref.~\cite{Tse23}, the critical temperatures of the associated BKT transitions for the 2D classical 6- and 8-state clock models
are calculated. The pseudo-critical temperatures are determined with a method that differs from the one used here.
Therefore, one needs to verify that the method employed in this study for calculating the pseudo-critical temperatures is valid for
a BKT transition.

\begin{figure}
       \includegraphics[width=0.6\textwidth]{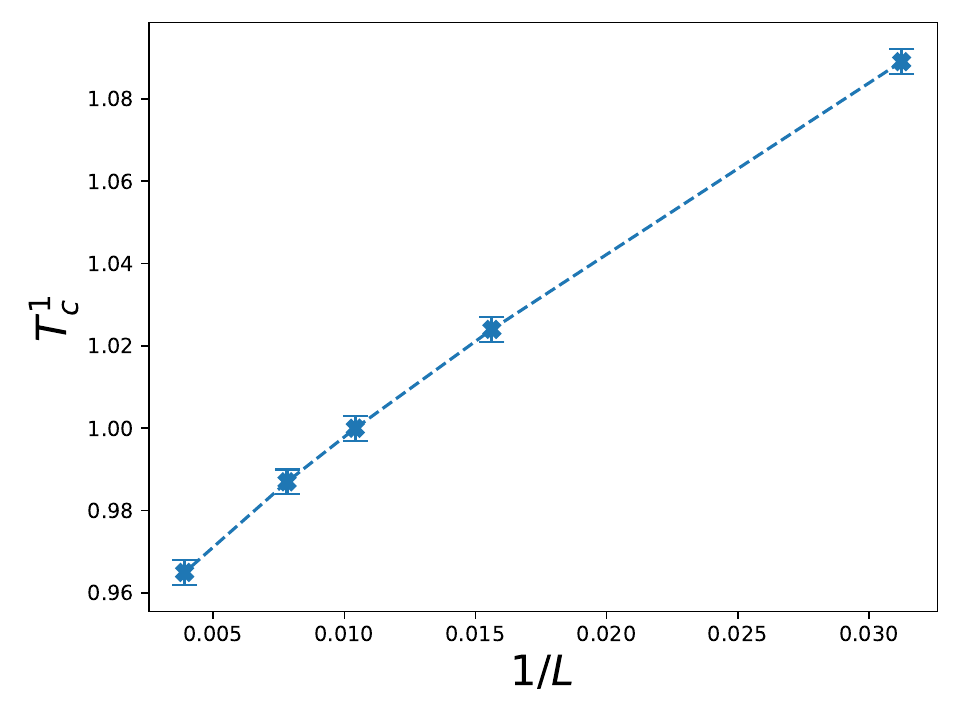}        
        \vskip-0.2cm
        \caption{Pseudo-critical temperatures $T_c^1(L)$ as a function
        of the linear box sizes $L$ for the 2D classical 6-state clock model.}
        \label{6clock}
\end{figure}

Using the data of Ref.~\cite{Tse23} as well as the idea of the intersection of $R$ and $(1+1/\sqrt{2})/2$, the pseudo-critical temperatures
$T_c^1(L)$ associated with the BKT transition from the pseudo-long-range order phase to the paramagnetic phase for the 6-state clock model
is shown in fig.~\ref{6clock}. With the data of fig.~\ref{6clock} and ansatz~(\ref{fit}), one arrives at $T_c^1 = 0.90(2)$ which agrees
quantitatively with the known result of $T_c^1 = 0.898(5)$ in the literature \cite{Sur19}.

\begin{figure}
       \includegraphics[width=0.6\textwidth]{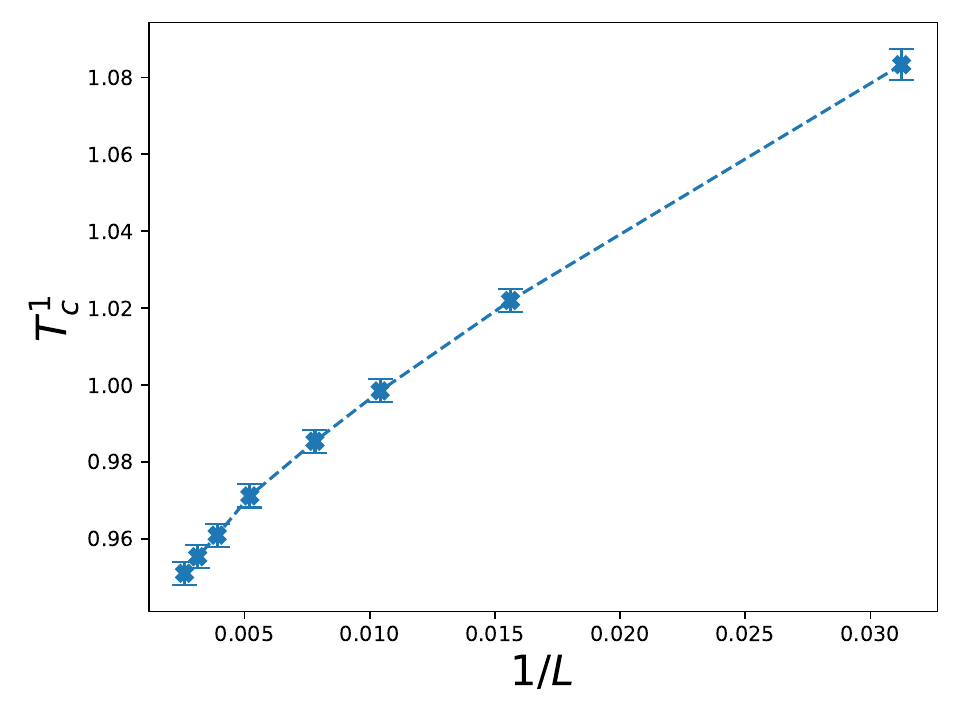}        
        \vskip-0.2cm
        \caption{Pseudo-critical critical temperatures $T_c^1(L)$ as a function
        of the linear box sizes $L$ for the 2D classical 8-state clock model.}
        \label{8clock}
\end{figure}

Similarly, with the data of the 8-state clock model from Ref.~\cite{Tse23} and the intersection method used here, one arrives at fig.~\ref{8clock}
regarding the associated $T_c^1(L)$. A fit of the data in fig.~\ref{8clock} and ansatz~(\ref{fit}) leads to $T_c^1 = 0.883(9)$.
The obtained $T_c^1 = 0.883(9)$ is in nice agreement with $T_c^1 = 0.8936(7)$ determined in Ref.~\cite{Tom022}.

The analysis associated with the 6- and 8-state clock models shown above implies the validity of our NN approach for calculating the critical
temperatures of BKT transitions.

In Ref.~\cite{Cam96}, by the method of strong coupling expansion, the inverse critical temperature $\beta_{\text{BKT,H}}$
of the 2D classical $XY$ model on the honeycomb lattice is found to be $\beta_{\text{BKT,H}} = 0.880$ (This leads to
$T_{\text{BKT,H}} = 1.1364$).
The $\beta_{\text{BKT,H}} = 0.880$ obtained in Ref.~\cite{Cam96} seems inconsistent with both ours and that of Ref.~\cite{Nie82}.
We would like to emphasize the fact that in Ref.~\cite{Cam96} the inverse critical temperature $\beta_{\text{BKT,S}}$ of
the 2D $XY$ model on the square lattice is determined to be 0.559 which is only about half of that calculated (by Monte Carlo simulations)
in Ref.~\cite{Has05} (The $\beta_{\text{BKT,S}}$ obtained in Ref.~\cite{Has05} is $\beta_{\text{BKT,S}} = 1.1199(1)$).
This implies it is likely that one needs to multiply the strong coupling expansion results of Ref.~\cite{Cam96} by a factor of 2 to obtain
the correct outcomes. If this is true, then the right $\beta_{\text{BKT,H}}$ based on the strong coupling expansion is given
by $\beta_{\text{BKT,H}} = 1.76$ with which one arrives at $T_{\text{BKT,H}}=0.56818$. The conjectured $T_{\text{BKT,H}}=0.56818$ is
in nice agreement with what's reached in this study. It is also worth noticing that while the inverse critical
temperature $\beta_{\text{BKT,T}}$ of
the 2D $XY$ model on the triangular lattice calculated from Ref.~\cite{Cam96} is 0.340, the $\beta_{\text{BKT,T}}$ estimated from
Monte Carlo simulations of Ref.~\cite{Sun22} is given by 0.676. Hence, it is of high possibility that the rule of
multiplication of factor 2 mentioned above is correct.

It should be pointed out that in Ref.~\cite{Nie82}, the model used for deriving
the critical temperatures of $O(N)$ vector models with $-2\le N \le 2$ is an
unphysical one, hence may not capture all the true features of the $O(N)$
universality class. 
The presented results in this investigation suggest that a refinement of
the related analytic calculation is required to better understand the
deviation of the numerical outcomes reached here from the theoretical
prediction.

\section*{Acknowledgement}
Partial support from the National Science and Technology Council (NSTC) of Taiwan is
acknowledged (NSTC 112-2112-M-003-016-). 
We thank Jhao-Hong Peng for providing us with some data so that we can verify the correctness of our Monte Carlo code.

\end{document}